\documentstyle[preprint,aps,prl]{revtex}  
\begin{document}
\title{On  Emery-Kivelson line and universality of Wilson ratio of
   spin anisotropic Kondo model}
\author{Jinwu  Ye}
\address{
Physics Laboratories ,       
Harvard  University, Cambridge, MA, 02138}
\maketitle
\begin{abstract}

 Yuval-Anderson's scaling analysis and Affleck-Ludwig's Conformal Field
 Theory approach are applied to the $ k $ channel {\em spin anisotropic}
 Kondo model. Detailed comparisons with the available Emery-Kivelson's Abelian
 Bosonization approaches are made. It is shown that the EK line exists
 for any $ k $, although it can be mapped to free fermions only when
 $ k=1 $ or $ 2 $. The Wilson ratio is universal if $ k=1 $ or $ 2 $,
 but {\em not} universal if $ k > 2 $.  The leading low temperature correction
 to the electron resistivity is {\em not} affected by the spin anisotropy for
 {\em any} $ k $. A new universal ratio for $ k>2 $ is proposed to compare with
 experiments.

\end{abstract}
\pacs{75.20.Hr, 75.30.Hx, 75.30.Mb}
\narrowtext

  In the general multichannel Kondo model, a magnetic impurity with spin 
 $ s $ couples to $ k $ degenerate bands of spin 1/2 conduction electrons
 by Heisenberg exchange interaction.
 When $ k \leq 2s $ (underscreening
  and complete-screening), the electron-impurity system can be described
 by a local Fermi liquid at very low temperatures. However, when $
 k > 2s $ (overscreening ), Nozi\'{e}res and Blandin \cite{blandin}
  showed that the low temperature physics, being controlled by a non-trivial
  zero temperature fixed point, is not described
  by Fermi liquid behavior. Bethe ansatz \cite{bethe}, Boundary conformal field
  theory (BCFT) \cite{affleck1}, Numerical renormalization group (NRG)
  \cite{affleck2}, Bosonization \cite{emery} etc. have been utilized 
  to investigate the nature of this 
   non-fermi liquid fixed point.
  By using BCFT approach, Affleck and Ludwig (AL)
  identified the non-trivial fixed point symmetry as Affine
 Kac-Moody algebra
 $ \widehat{SU}_{k}(2) \times \widehat{SU}_{2}(k) \times U(1) $.
  Near the fixed point, they classified all the possible perturbations 
  according to the representation theory of the underlying
  KM algebra at the fixed point.
  AL only considered isotropic case, although they showed \cite{affleck2} that
  spin anisotropy is always irrelevant for $ s=1/2 $ {\em  in the sense 
    that no relevant operators
   will appear by allowing spin anisotropy }. BCFT is very elegant and
  applicable for any channel. The symmetry is also explicitly demonstrated
   in this approach. But the relation between the boundary operators of BCFT
    and the original scaling variables is not transparent and it is hard
  to apply CFT near the weak coupling {\em fixed line}.

    Emery and Kivelson (EK)~\cite{emery,emery2},
 using Abelian Bosonization,
  found an alternative solution to $ k=2, s=1/2 $ anisotropic Kondo model.
   By using a canonical transformation $ U= e^{i S_{z} \Phi_{s}(0) } $
    and refermionization,
  they located a exactly solvable line (EK line) which
   is analogous to the Toulouse line \cite{toul}  for the
   ordinary single channel Kondo model.  They also 
   found {\em one } leading irrelevant operator $ S_{z} \partial_{x} \Phi_{s} $
    away from the EK line. By doing
   perturbative calculations around the EK line with this operator,
   they recovered the generic low temperature 
   behaviors of the impurity specific heat and susceptibility.
   Sengupta and Georges \cite{georges}, using EK's method,
   calculated the Wilson ratio independently.
   Later EK's approach was extended to the 4 channel case \cite{gogolin} and
   applied to the electron assisted tunneling of a heavy particle
    between two sites in a metal \cite{fisher}.
   EK's approach can be applied easily, but it is limited to special values 
   of $k $. The symmetry is hidden and there
   is no systematic classification of all the possible operators in this
   approach.
   
  The two channel Kondo model has been argued to describe
  uranium based heavy fermion systems and tunneling of a non-magnetic impurity
  between two sites in a metal \cite{cox}.  $ k=1, 2 $ and $ k=3 $
 {\em magnetic } Kondo models have also been proposed to interpret $ Ce^{3+} $
  based alloys \cite{kim}. It is believed that
  the general spin anisotropy of the form
 in Eq.\ref{kondox} is ubiquitous in all the experimental systems
 mentioned above, therefore it is important to investigate its effects
 in detail.
 
  In this paper, fixing the impurity spin $ s=1/2 $, we extend the scaling
  analysis of YA near the weak coupling fixed line to the
   {\em multi-channel} Kondo model
    and AL's approach near the intermediate coupling fixed point to the
   {\em  spin anisotropic} Kondo
   model. We also make detailed comparisons among YA's, AL's and EK's approaches.
    We find 1-1 mapping between the boundary operators in CFT and those in
    EK's solution, therefore establish the relation between the
    scaling parameters in the two approaches.  We define the
   EK line as a special line in parameter space
    where the impurity part of Kondo system completely decouples
   from the uniform external magnetic field, hence $ \chi_{imp} \equiv 0 $
     \cite{tsvelik}. The EK line defined in this way
    coincides with the conventional EK line
    if $ k=2 $ and the decoupling line \cite{emery2} if $ k=1 $.
    We show that the EK line is a natural property for {\em any $ k $}
   channel Kondo model, although this line can be mapped to free theory
   only when $ k= 1, 2 $ \cite{free}. The reason that only {\em one } leading 
   irrelevant operator was found in EK's approach to 2 (1) channel
   Kondo model is due to first order (second order)
    KM null states \cite{witten}, 
    therefore the Wilson ratio is universal for the {\em general} spin anisotropic
    2 (1) channel-Kondo model Eq.\ref{kondox}.  However, the Wilson ratio is not universal
   for $ k>2 $ due to {\em two or more } leading irrelevant operators with the 
   same scaling dimension. It is shown that the {\em general} spin
    anisotropy does {\em not} affect
    the leading low temperature correction to the electron resistivity
    for {\em any k}.
    A new universal ratio for $ k>2 $ is proposed to compare with
    experiments.

   The three-dimensional Kondo problem can be mapped to a one dimensional
  problem \cite{affleck1}. Following closely the notation of AL,
   we write the following
  $ k $ channel {\em general} spin anisotropic Kondo Hamiltonian
   in a uniform magnetic field 
 \begin{equation}
 H=\frac{v_{F}}{2 \pi} (\int^{\infty}_{-\infty} dx i
   \psi^{\dagger}_{i \alpha L}(x) \frac{d \psi_{i \alpha L}(x)}{dx}
   +2 \pi \sum_{a,b=x,y,z} \lambda^{ab}_{K} J^{a}_{L}(0)  S^{b} )
   +h ( \int dx J^{z}_{L}(x) + S^{z} )
\label{kondox}
 \end{equation}
   where $ J^{a}_{L}(x) =\frac{1}{2} \psi^{\dagger}_{i \alpha L}
   \sigma^{a}_{\alpha \beta} \psi_{i \beta L}(x) $ is the spin
  current of the conduction electrons,
  $ S^{a} $ $  (a=x,y,z) $  is the impurity spin \cite{lande}.
   In most of this paper, we limit our discussions 
   to XXZ ( $ U(1) \times Z_{2} \sim O(2) $ ) case.

{\em Weak coupling analysis}:
   We extend the YA's scaling equations
 \cite{anderson} near the weak coupling fixed line to the $ k $
 channel case \cite{jinwu3}:

\begin{eqnarray}
 \frac{ d Q}{d l} & = & -\frac{1}{2} ( k Q-1) \lambda^{2} + \cdots
  \nonumber \\
 \frac{ d \lambda}{ d l} & = & [ 1- (k-1) Q^{2} -( Q-1)^{2} ] \lambda  +\cdots 
\end{eqnarray}

   where $ Q= \frac{ \delta }{\pi} , \delta= 2 \delta_{+},
   \delta_{+} =- \delta_{-} = \tan^{-1} (\frac{\pi \lambda_{z}}{4}) $.

  Defining $ Q= \frac{1}{k} + q $. It is easy to see that,
  near the fixed point $ (q_{0}, 0 ) $, there is only {\em one } relevant
 direction. The Kondo scale is given by $ T_{K} \sim D
  \lambda ^{k/( 1-(kq_{0})^{2})} $. The special line $ q=0 $ where phase 
  shift $ \delta_{+}= \frac{\pi}{ 2k } $ will be discussed later.

{\em Intermediate coupling analysis}:
   In order to locate the intermediate-coupling fixed point, we set
 $\lambda_{K}^{ab}=\lambda_{K} \delta^{ab}, h=0 $;
  then the Hamiltonian Eq.\ref{kondox}
   has global $ SU(2) \times SU(k) \times U(1) $  symmetry.
 Fourier transforming  Eq.\ref{kondox}
 on a finite line segment $-l \leq x \leq l $ of length
 $2l $, the spin part of the Hamiltonian which contains the Kondo interaction
 becomes
\begin{equation}
H_{s} = \frac{v_{F} \pi}{l} ( \sum_{n=-\infty}^{\infty} \frac{1}{2+k}
 : \vec{J}_{-n} \cdot \vec{J}_{n} : + \lambda_{K} \sum_{n=-\infty}
^{\infty} \vec{J}_{n} \cdot \vec{S} )
\label{kondop}
\end{equation}
 where $ \vec{J}_{n}'s $  are the Fourier modes of the spin current operator:
 $\vec{J}_{n} =\frac{1}{2 \pi} \int_{-l}^{l} dx e^{in\pi x/l} \vec{J}_{L}(x)$.
 They obey the $ SU(2) $ level $ k $ KM commutation relations
\begin{equation}
[ J^{a}_{n} , J^{b}_{m} ] = i\epsilon^{abc} J^{c}_{n+m} + \frac{k}{2}
n \delta^{ab} \delta_{n+m,0}
\label{KM}
\end{equation}
    
  When $\lambda_{K}=\lambda^{*}_{K}=\frac{2}{2+k} $ to be identified as
 the intermediate coupling fixed point, the global symmetry is enlarged to a 
 local KM symmetry
 $ \widehat{SU}_{k}(2) \times \widehat{SU}_{2}(k) \times U(1) $,
 the interacting spin Hamiltonian of Eq.\ref{kondop} is exactly
  the same as free spin Hamiltonian when written
 in terms of the shifted current operators $ \vec{ \cal{J}}_{n}
 =\vec{J}_{n}+ \vec{S} $ which obey the same KM algebra Eq.\ref{KM}
  as the free operators.
       
   Near the fixed point, the Hamiltonian can be written as the
  fixed point Hamiltonian plus possible perturbations:
\begin{equation}
  H= H^{*} + \sum_{i} \lambda_{i} O_{i}(0)
\end{equation}
    
   We can classify all the possible perturbations $ {O_{i}} $ in the 
  physical problem according to the representation theory of the underlying
  KM algebra at the fixed point. The possible operators which can take
  us  away from the fixed point should also respect the global symmetry
  of the problem.
    According to the fusion rule \cite{affleck1},
    for the overscreened case, the spin-1 ( adjoint)
  representation is always allowed, its first order KM descendants have
  nine Cartesian tensor operators of rank two
   $ {\cal J}_{-1}^{a} \phi^{b} $ which can be decomposed into the irreducible
  spherical tensors under $ SU(2) $ with angular momentum $ j=0, 1, 2$.
\begin{eqnarray}
  T^{0}_{0} &=& \vec{{\cal J}}_{-1} \cdot \vec{\phi}  \nonumber\\
  T^{1}_{a} &=& L_{-1} \phi^{a}   \nonumber \\
  T^{2}_{\pm 2} &=&  {\cal J}_{-1}^{1} \phi^{1}- {\cal J}_{-1}^{2} \phi^{2}
         \pm i( {\cal J}_{-1}^{1} \phi^{2} +{\cal J}_{-1}^{2} \phi^{1} ) 
     \nonumber  \\
  T^{2}_{\pm 1} &=& \mp [ {\cal J}_{-1}^{1} \phi^{3}+ {\cal J}_{-1}^{3} \phi^{1}
         \pm i( {\cal J}_{-1}^{2} \phi^{3} +{\cal J}_{-1}^{3} \phi ^{2} )]  \nonumber  \\
  T^{2}_{0} &=& {\cal J}_{-1}^{3} \phi^{3}-\frac{1}{2} ({\cal J}_{-1}^{1} \phi^{1} +
   {\cal J}_{-1}^{2} \phi^{2} )      
\label{tensor}
\end{eqnarray}
    where $ L_{-1} $ is the Virasoro lowering operator.

    In the following, we will discuss
  $ k >2, k=2 $ and $k =1 $  respectively.

       { \em  $ k >2 $  case }:
     For the isotropic  $ SU(2) $ case, AL identified  $ T^{0}_{0} =\vec{{\cal J}}_{-1} \cdot \vec{\phi} $
   as the only leading irrelevant operator
   whose scaling dimension is $ 1+ \Delta $ ($ \Delta = \frac{2}
  {2+k} $ is the scaling dimension of $ \vec{ \phi} $),
   the corresponding coupling constant 
  $\lambda_{0}$ has R. G. eigenvalue
  $- \Delta  < 0  $, therefore is irrelevant.

   For $ U(1) \times Z_{2} $ case, the second operator $ T^{2}_{0} $
  is also allowed by  all the symmetry of the model. We regroup the two
  boundary operators as :
  \begin{eqnarray}
    O & = & \lambda_{0} T^{0}_{0} + \lambda_{2} T^{2}_{0} =
         \alpha_{1} O_{1} + \alpha_{2} O_{2}    \nonumber  \\
    O_{1}  & = &  (\frac{k}{2}-1) T^{0}_{0} -(\frac{k}{2} +2) T^{2}_{0}
        \nonumber   \\
    O_{2} & = & T^{0}_{0}+2 T^{2}_{0}=3 {\cal J}_{-1}^{3} \phi^{3} 
  \end{eqnarray} 

     We can easily extend  Eq.(3.29) of Ref.\cite{affleck1} to
\begin{eqnarray}
  exp(-\beta f_{imp}(T,\lambda_{0},\lambda_{2},h ))  & =  &
   exp(-\beta f_{imp}(T,0,0,0))
   \langle exp\{ h \frac{3k}{2} \alpha_{2} \int^{\beta/2}_{-\beta/2}
   d \tau \phi^{3}(\tau)               \nonumber   \\
    &  & ~~~~ + \alpha_{1} \int^{\beta/2}_{-\beta/2} 
   d \tau O_{1}(\tau) +
   \alpha_{2} \int^{\beta/2}_{-\beta/2} d \tau O_{2}(\tau) \} \rangle_{T}
\label{exten}
\end{eqnarray}
   
  Observing $ h $ acquires no anomalous dimension to any loops 
  and applying finite size scaling lead to
  the following scaling form of the impurity free energy at low temperature
 \begin{eqnarray}
    -\delta f_{imp}(T, \alpha_{1}, \alpha_{2},h) & = &  
    T Q_{imp}(h/T, \alpha_{1} T^{\Delta}, \alpha_{2} T^{\Delta})
           \nonumber  \\
       =  T \{ A [ \frac{9}{8} k(k-2)(\frac{k}{2} +2 ) \alpha_{1}^{2} & + &
             \frac{9}{2} k \alpha_{2}^{2} ] T^{2 \Delta}  + 
     G( h \frac{3k}{2} \alpha_{2} T ^{\Delta -1} ) \} (1 + O(T) )
\label{ascaling}
 \end{eqnarray}
   where $ \delta f_{imp} (T, \alpha_{1}, \alpha_{2}, h) =
  f_{imp} (T, \alpha_{1}, \alpha_{2}, h)
   +T \log \sqrt{2} $ and G takes the following  asymptotic form
\begin{equation}
   G(s)= \left \{ \begin{array}{ll}
                   \sim s^{2}    &  s \rightarrow 0  \\
                   \sim s^{1/(1-\Delta)}  &   s \rightarrow \infty
                   \end{array}   \right.  
\end{equation}

    From the above scaling equation, it is easy to see 
\begin{eqnarray}
     C_{imp} & \sim & [ \frac{9}{8} k(k-2)(\frac{k}{2} +2 ) \alpha_{1}^{2}+
             \frac{9}{2} k \alpha_{2}^{2} ] T^{2 \Delta}     \nonumber  \\
    \chi_{imp} & \sim &  
     ( \frac{3k}{2} \alpha_{2})^{2} T ^{2 \Delta -1}  
\label{thermal}
\end{eqnarray}

     The Wilson ratio $ R= \frac{ T \chi_{imp} } { C_{imp} } $
    is not a universal constant {\em  in contrast to} the isotropic case.
  The $ h $ dependent part of the zero temperature impurity free energy
  is given by :
\begin{equation} 
   f_{imp}(0,\alpha_{1}, \alpha_{2},h) 
    \sim (\frac{3k}{2} \alpha_{2} h)^{1/(1-\Delta)} 
  ~~~~  h \rightarrow 0 
\end{equation}

  Therefore the impurity susceptibility is given by $ \chi_{imp}
   =\frac{m_{imp}}{h} \sim h ^{\frac{2 \Delta -1}{1-\Delta}} $  at $ T=0 $.

    If $ \alpha_{2} =0 $, the impurity part of the system decouples
   from the external magnetic field, $\chi_{imp} $ vanishes to the
  leading order. This is exactly the feature of the EK line in the
    $ k >2 $ case.
    Extending EK's approach to the four channel Kondo model,
    Fabrizio and Gogolin \cite{gogolin} located the EK line which,
   unlike the $ k=2 $ case, {\em cannot}  be transformed to a Hamiltonian
    for free fermions by refermionization. They managed to find two orthogonal
    operators with scaling dimension 4/3, one contributes only to 
    $C_{imp} $, another contributes to both $ C_{imp} $ and
    $ \chi_{imp} $. Their results are totally consistent with ours.
    By symmetry we can identify $ O_{2}= {\cal J}_{-1}^{3} \phi^{3} $ with 
    $  S_{z} \partial_{x} \Phi_{s} $  found by them.

      If $ \alpha_{1} =\alpha_{2} =0 $, the next leading irrelevant
   operators belong to the conformal tower of identity operator with scaling
   dimension 2.  Similar to Eq.\ref{tensor}, they can be classified as
\begin{eqnarray}
  Q^{0}_{0} &=& {\cal J}_{-1}^{a} {\cal J}_{-1}^{a} I     \nonumber\\
  Q^{1}_{a} &=& {\cal J}_{-2}^{a} I = \partial {\cal J}^{a}     \nonumber \\
  Q^{2}_{\pm 2} &=&  [ {\cal J}_{-1}^{1} {\cal J}_{-1}^{1}- {\cal J}_{-1}^{2} {\cal J}_{-1}^{2}
         \pm i( {\cal J}_{-1}^{1} {\cal J}_{-1}^{2} +{\cal J}_{-1}^{2} {\cal J}_{-1}^{1} )] I \nonumber  \\
  Q^{2}_{\pm 1} &=& \mp [ {\cal J}_{-1}^{1} {\cal J}_{-1}^{3}+ {\cal J}_{-1}^{3} {\cal J}_{-1}^{1}
         \pm i( {\cal J}_{-1}^{2} {\cal J}_{-1}^{3} +{\cal J}_{-1}^{3} {\cal J}_{-1}^{2} )] I  \nonumber  \\
  Q^{2}_{0} &=& [ {\cal J}_{-1}^{3} {\cal J}_{-1}^{3}-\frac{1}{2} ({\cal J}_{-1}^{1} {\cal J}_{-1}^{1} +
   {\cal J}_{-1}^{2} {\cal J}_{-1}^{2} ) ] I      
\label{tensor2}
\end{eqnarray}

   We regroup the two boundary operators $ Q^{0}_{0}, Q^{2}_{0} $ as :
  \begin{eqnarray}
    P & = & \beta_{1} P_{1} + \beta_{2} P_{2}    \nonumber  \\
    P_{1}  & = &  (k-1) Q^{0}_{0} -(k +2) Q^{2}_{0}
        \nonumber   \\
    P_{2} & = & Q^{0}_{0}+2 Q^{2}_{0}=3 {\cal J}_{-1}^{3} {\cal J}_{-1}^{3} I
  \label{tower}
  \end{eqnarray}
 
      Although the operators $ P_{1} $ and $ P_{2} $ are Virasoro descendants,
    therefore have non-vanishing expectation values at finite temperature, 
   it still can be shown that \cite{jinwu3}
   $ P_{1} $, running along the EK line, contributes only to
    $ C_{imp} \sim T $; $ P_{2} $, running away
   from the EK line, contributes to both $ C_{imp} \sim T $ and $ \chi_{imp}
   \sim const. $

       We can continue this process to any level of KM towers and locate
   the EK line where $ \alpha_{2} =\beta_{2}= \cdots=0 $,
    hence $ \chi_{imp} \equiv 0 $. The coefficients $ \alpha_{1},
     \alpha_{2}, \beta_{1}, \beta_{2} \cdots $ all depend on $ q_{0} $
    where we start at high energy scale. The precise relation may be obtained
  by Bethz ansatz solution \cite{tsvelik}. $ \chi_{imp} $ has also been shown
   \cite{yuval} to vanish at the special line
   $ q_{0}=0 $ found at the weak coupling analysis, therefore if $ q_{0} =0 $,
   then $ \alpha_{2} =\beta_{2} = \cdots =0 $.

   If we apply a local magnetic field to the system, the 
   boundary operator \cite{affleck3}
   $ h S^{z} = h(\phi^{3} + {\cal J}^{3} + L_{-1} \phi^{3} + \cdots) $
  is added to the fixed point Hamiltonian. Unlike in the uniform field,
   this boundary operator is totally {\em independent } of other boundary
   operators, therefore we always have $ \chi^{l}_{imp} \sim h^{ \frac{2 \Delta-1}
   {1-\Delta}} $  at $ T=0  $, {\em non-vanishing} even at the EK line.

  In the isotropic case, as shown by AL~\cite{affleck3,resitivity},
  the {\em first } order perturbation in $ T^{0}_{0} $ gives the
  leading low temperature correction to the electron resistivity.
  Because $ T^{2}_{q} (q=0, \pm1, \pm2) $ carry spin 2, the first order
   perturbations
  in these operators {\em vanish}, therefore the leading low temperature correction
   to the electron resistivity is {\em not} affected by the {\em general}
   spin anisotropy in Eq.\ref{kondox} 
\begin{equation}
  \rho(T) \sim \rho(0) (1 + \lambda_{0R} T^{ \Delta} )
       \sim \rho(0) (1 +[( \frac{k}{2}-1)\alpha_{1} + \alpha_{2}] T^{\Delta})
\label{resitivity}
\end{equation}

     Eqs.\ref{thermal}, \ref{resitivity} can be used to get rid of the
 two independent parameters $ \alpha_{1}, \alpha_{2} $, therefore a new
  universal ratio can be formed. Other transport
 properties like thermopower and thermal conductivity can also be calculated,
  more universal ratios can be formed \cite{read}.

   The predictions of this section may be tested by
   the possible experimental realization of the 3 channel Kondo model
   proposed in Ref. \cite{kim}.

    {\em  $ k=2 $ case} : When $ k=2 $, the coefficient of
  $ \alpha_{1} $ in Eq.\ref{ascaling}
  vanishes. The underlying reason for this is that the pure KM null
   states appear
  in the first order descendant for the $ j=1 $ representation.
 Generally, for the spin j representation of $ \widehat{ SU}_{k}(2) $,
  the pure KM null states
  \cite{witten} first appear at $( k-2j+1) $'s order:
  $ ( {\cal J}_{-1}^{+}  )^{k-2j+1} |(j),j > =0 $  ( $ {\cal J}_{-1}^{\pm}=
   {\cal J}_{-1}^{1}  \pm i {\cal J}_{-1}^{2} $ ). For $ j=1 $, it becomes 
  $ ({\cal J}_{-1}^{+} )^{k-1} | (1),1 > =0 $. If $ k > 2 $, the null states first
 appear at least at order 2, so the nine operators defined in Eq.\ref{tensor}
  are all independent of each other. However, for $
  k=2 $, there are 6 constraints:
 $ {\cal J}_{-1}^{\pm}|1,\pm1> =0 $ plus cyclic
  permutations in x and y axis. It turns out the 6 constraints only
  give  5 independent equations: $ T^{2}_{q} =0, q=0,\pm 1, \pm 2 $, therefore
  there is only one leading irrelevant operator $ T^{0}_{0} $ 
  even for $ U(1) \times Z_{2} $ case . Actually we can 
  make an even stronger statement:
 for the {\em general} anisotropic 2-channel Kondo model Eq.\ref{kondox},
   there is only {\em one } leading irrelevant operator
  $ T^{0}_{0} $.  The EK line is given by $ \alpha_{2}= \beta_{2}=0 $
   which corresponds to $ \delta_{+} =\frac{\pi}{4} $
  in the weak coupling analysis. The leading
  irrelevant operator along this line
   is  $ P_{1} =-3[{\cal J}_{-1}^{3} {\cal J}_{-1}^{3}-({\cal J}_{-1}^{1} {\cal J}_{-1}^{1} +
   {\cal J}_{-1}^{2} {\cal J}_{-1}^{2} ) ] I  $ of Eq.\ref{tower},     
   the {\em first} order perturbation in $ P_{1} $ gives $ C_{imp} \sim T $. R. G.
   analysis of EK's solution also show that the leading irrelevant operator {\em along} the EK
   line has scaling dimension 2 \cite{jinwu3}.  
   However, away from the EK line, $ T^{0}_{0}= 3 {\cal J}_{-1}^{3} \phi^{3} $
   which is Virasoro primary
   is the only leading irrelevant operator with scaling dimension 3/2. 
    By  symmetry we can identify $ T^{0}_{0}= 3 {\cal J}_{-1}^{3}
    \phi^{3} $ with 
    $ S_{z} \partial_{x} \Phi_{s} $ found by EK, therefore $ \lambda_{0} \sim
     \lambda_{z}- 2 \pi v_{F} $. This implies that
    $\lambda_{0} $ {\em changes sign} as it passes through the fixed point,
     therefore confirms the conjecture by AL \cite{resitivity} that the
    resistivity shows very different behaviors on the two sides of
     the fixed point.  We can also identify $ 
    P_{2} =3 {\cal J}_{-1}^{3} {\cal J}_{-1}^{3} I $ as 
    $ : ( \partial_{x} \Phi_{s} )^{2} :$, therefore
    $ \beta_{2} \sim (\lambda_{z}-2 \pi v_{F} )^{2} $. 

    {\em  $ k=1 $ case} : Only the $ j=0 $ representation is allowed
  by the fusion rule in the ordinary 1-channel Kondo model \cite{affleck1}.
   For the spin 0 representation of $ \widehat{ SU}_{k}(2) $,
  the pure KM null states first appear at $( k+1) $'s order:
  $ ( {\cal J}_{-1}^{+}  )^{k+1} |(0),0 > =0 $. For $ k > 1 $, the null states first
 appear at least at order 3, so the nine operators defined in Eq.\ref{tensor2}
  are all independent of each other. However, for $ k=1 $, there are 6 constraints:
 $ ( {\cal J}_{-1}^{\pm} )^{2}| (0),0> =0 $ plus cyclic
  permutations in x and y axis, they 
  lead to $ Q^{2}_{q} =0, q=0, \pm1, \pm2 $. There is only
  {\em one} leading irrelevant operator $ Q^{0}_{0} $ 
  for the {\em general} anisotropic 1-channel Kondo model Eq.\ref{kondox}.
  The {\em first } order perturbation
  in $ Q^{0}_{0} $ \cite{jinwu3} leads to the generic fermi-liquid behaviors
   : $ \chi_{imp} =const, C_{imp} \sim T, R \equiv 2 $. 

   EK's solution of the $ O(2) $ one channel Kondo model \cite{emery2} found
  a decoupling line which corresponds to $ \delta_{+} = \frac{\pi}{2} $
   in the weak coupling analysis. R. G. analysis \cite{jinwu3} near this line
   show that there is only {\em one} dimension 2 operator away from
   the fixed point.
   We can identify $ Q^{0}_{0} =3 {\cal J}_{-1}^{3} {\cal J}_{-1}^{3} I $
   as $ : ( \partial_{x} \Phi_{s} )^{2} : $, therefore $ \lambda_{0} \sim
    (\lambda_{z} - 4 \pi v_{F} )^{2} $. This implies $\lambda_{0} $
   is always {\em positive}, {\em in contrast to} the 2 channel case.

We thank D. S. Fisher, B. Halperin, D. Cox, N. Read, S. Sachdev, 
  A. L. Moustakas, D. Anselmi and A. Tsvelik
  for helpful discussions. We have benefited from the discussions
  on CFT with S. Sethi and E. Zaslow.
This research was supported by NSF Grants Nos. DMR 9106237 and DMR9400396.

  After the completion of this paper, I was informed by A. Georges
  \cite{georges2} that he and A. M. Sengupta interpreted
  the canonical transformation in EK's solution as a boundary condition
  changing operator \cite{xray}. I thank him for pointing out this
  connection to me. Maldocena and Ludwig \cite{ludwig} discussed the connection
  between CFT and Abelian Bosonization, but with different emphasize
  than this paper.


\begin{references}
\bibitem{free}  $ \widehat{SU}_{k}(2) $ has free fermion representation if $ k=2 $, vertex operator construction if $ k=1 $.
\bibitem{lande} We assume the electron and the impurity have the same
       Land\'{e} factors.
\end{references}
\end{document}